\documentstyle[sprocl]{article}

\input{psfig}

\def\Journal#1#2#3#4{{#1} {\bf #2}, #3 (#4)}

\def\APJ{{\em Astrophys. J.}}
\def\APJL{{\em Astrophys. J. Lett.}}
\def\APJS{{\em Astrophys. J. Suppl.}}
\def\MNRAS{{\em Mon. Not. Roy. Astron. Soc}} 
\def\MSAIT{{\em Mem. Soc. Astron. It.}} 
\def\AJ{{\em Astron. J.}}
\def\AAP{{\em Astron. \& Astrophys.}}

\def\be{\begin{equation}}
\def\ee{\end{equation}}
\def\bea{\begin{eqnarray}}
\def\eea{\end{eqnarray}}

\newcommand{\kms}{$km~s^{-1}$ }
\newcommand{\eg}{{\it e.g., }} 
\newcommand{\ie}{{\it i.e.~}} 
 
\newcommand{\etal}{{\it et al.}} 
\newcommand{\msun}{$M_\odot$ }

\begin{document}

\title{SUPERNOVAE} 

\author{N. PANAGIA} 

\address{ESA-STScI, 3700 San Martin Drive, Baltimore, \\ MD 21218, USA
\\E-mail: panagia@stsci.edu} 


\maketitle\abstracts{ 
The properties of supernovae (SNe) are reviewed. It is shown that the
observed characteristics of the morphological classes of SNe (types
Ia, Ib/c, II) can be explained in terms of two basic explosion 
mechanisms, \ie core collapse of massive stars and thermonuclear
explosion of low/moderate mass stars. 
The study of SNe can provide valuable insight in the late phases
of the evolution of their progenitors and, therefore, can constrain
the theory of stellar evolution quite tightly. 
Also, bright SNe of type Ia can be used to probe the Universe up to 
high redshifts, allowing us to measure cosmological constants and to 
gain valuable information on the formation and evolution of galaxies.
The case of SN~1987A is discussed in some detail: it appears that in
this event a number of predictions were astonishingly well verified
but still many aspects were completely at variance with ``common
wisdom" expectations, indicating that the SN phenomenon is still not
fully understood. 
}

\section{Introduction}

As their name indicates, {\it supernovae} (SNe) are discovered in the
sky as ``new stars" ({\it -novae}) of exceptionally high brightness
({\it super-}). The fact that SNe are formidable explosions completely
different from, and vastly more energetic than classical {\it novae}
\footnote{Novae are produced by sudden nuclear ignition of a very thin
layer of hydrogen near the surface of a degenerate star accreting
matter from a binary companion.}, was first recognized by  Baade and
Zwicky (1934). They noticed that novae during explosion become no
brighter than about 1 million times (\ie 15 mag) they are in a
quiescent phase. Therefore, any historical event in our Galaxy that
had reached a magnitude as bright as 0 or brighter but is not
detectable at present, had  to belong to a separate class of
intrinsically brighter objects.  And indeed, the distribution of
observed magnitudes of explosive events detected in galaxies of the
Local Group indicated the presence of two peaks, one at the expected
brightness of classical novae and another at more than thousand times
brighter luminosities. 

Supernovae represent the explosive death of both low mass stars (type
Ia) and moderate and high mass stars (types Ib/c and II). They are
extremely bright, (roughly 10$^9~L_\odot$\footnote{In Astronomy the
symbol $\odot$ is used to denote the Sun. Thus, $L_\odot =
3.8\times10^{33}~ erg~s^{-1}$ is the solar luminosity and $M_\odot =
2.0\times10^{33}~g$ is the solar mass.} rivalling, for a few days, the
combined light of the entire host galaxy. In all cases, a SN explosion
injects highly metal-enriched material (at least 1 $M_\odot$) and a
conspicuous amount of kinetic energy (about 10$^{51}$ ergs) into the
surrounding medium (see Section 2.1). In addition, the blast waves
from SN explosions produce powerful sources of radio and X-ray
emission -- supernova remnants -- that can be seen and studied many
thousands of years after the event (see Section 2.2). Therefore, it is
clear that SN explosions are crucial events that determine most of the
aspects of the evolution of galaxies, \ie most of the visible
Universe.

Some SNe in our Milky Way galaxy have been close enough to be visible
to the naked eye, and records of their occurrence can be found in
ancient annals. In particular, during the past 2000 years 9 such
events have been recorded. A few of these events were very bright. The
supernova of 1006 AD, for example, was about 1/10 as bright as the
full moon! The last supernova to be seen in our Galaxy was discovered
in 1604 by the famous astronomer Kepler. On the basis of these
historical  records one may infer that the average rate of SN
explosions in the Galaxy be of the order of 5 per millennium. However,
one has to allow for the fact that most SNe are either too far or are
too obscured by dark dust clouds of the galactic disk to be visible.
Actually, one can estimate that only about 10\% have been close enough
and bright enough to be detectable by naked eye. Therefore, a more
realistic SN explosion rate for our Galaxy is about one every twenty
years (see also Section 2.3). 

Being so bright, SNe are ideal probes of the distant Universe.  And
indeed studies of SNIa up to redshifts $\sim1.2$ have allowed us to
explicitely measure both the local expansion rate of the Universe and
other cosmological parameters (see Section 2.4). 
The brightest supernova discovered in the last three centuries is
supernova 1987A in the Large Magellanic Cloud, a small satellite galaxy
to the Milky Way. Section 3 is devoted to it.

\section{Properties of Supernovae} 

\subsection{Supernova Types}

Morphologically, Supernovae are distinguished into two main classes, Type
I and Type II according to the main criterion of whether
their spectra (thus, their ejecta) contain Hydrogen (Type II)
or no Hydrogen (Type I). 

Type II SNe are produced by the core collapse of massive stars, say, more
massive than 8 $M_\odot$ and at least as massive as 20 $M_\odot$
(SN~1987A) or even 30 or more $M_\odot$ (SN~1986J).  Thus, the lifetime
of a SNII progenitor is shorter than about 100 million years (and can be
as short as a FEW million years). Therefore, SNII can be found only in
galaxies that are either just formed or that have efficient, ongoing
star formation, such as spiral and irregular galaxies.

\begin{table}[t]   
\caption{GENERAL PROPERTIES OF SUPERNOVAE\label{tab:proper}}
\vspace{0.2cm}
\begin{center}
\footnotesize
\begin{tabular}{|c|ccc|}
\hline
& & & \\
Type               &   Ia          &   Ib/c         & II		\\
& & & \\
\hline
& & & \\
Hydrogen           &    NO         &    NO          &  YES		\\
& & & \\
Optical            &  Metal lines  &  Metal lines   & P Cyg lines	\\
Spectrum           & deep 6150 \AA & no 6150 \AA    & Balmer series	\\
& & & \\
Absolute           & $\sim4\times~10^9~L_\odot$   & $\sim10^9~L_\odot$ 	&   
		$\sim10^9~L_\odot$	\\
Luminosity         & small dispersion & small disp.? & large disp.	\\
at max light       & standard candles & 	    & 			\\
& & & \\
Optical            & homogeneous   &   rather       & heterogeneous	\\
Light Curve        &               & homogeneous    &			\\
& & & \\
UV spectrum        & very weak     &    weak        &  strong		\\
& & & \\
Radio              & no detection  &   strong       &  strong		\\
Emission           &               & fast decay     & slow decay	\\
& & & \\
Location           & all galaxies  &  spirals       &  spirals \&	\\
                   &               &                & irregulars	\\
& & & \\
Stellar            &    old        &   young        &   young		\\
Population         &               &                &                   \\
& & & \\
Progenitors        & white dwarfs  & moderately     & massive stars 	\\
                 & in binary systems  & massive stars &			\\
& & & \\
\hline
\end{tabular}
\end{center}
\end{table}

The class of Type I supernovae has been recognized (\eg Panagia 1985) to
consist of two subclasses, Type Ia and Type Ib/c that, although sharing
the common absence of Hydrogen, are widely apart in other properties
and, especially, in their origins. The spectroscopic criterion to
discern the two subclasses from each other is the presence (Ia) or
absence (Ib/c)\footnote{They are classified Ib if strong He lines are
present in their spectra, and Ic otherwise.} of a strong Si$^+$ 6150\AA~ 
absorption feature which is prominent in their early epoch spectra. The
astrophysical difference between Type Ia and Ib/c SNe is that the former
are found in all type of galaxies, from ellipticals through spirals to
irregulars, whereas the latter are found exclusively in spiral galaxies,
mostly associated with spiral arms and frequently in the vicinities of
large ionized nebulae (giant HII regions). These characteristics
indicate that SNIb/c are the end result of a relatively young population
of stars (ages less than 100 million years) while SNIa progenitors must
be stellar systems that have considerably longer lifetimes, of the order
of 10$^9$ years or more. 

The progenitors of SNIa are believed to be stars that would not 
produce a SN explosion if they were single stars but that end up
exploding because, after reaching the white dwarf stage, they accrete
enough mass from a binary companion to exceed the Chandrasekhar mass,
and ignite explosive nucleosynthesis in their cores. This process of
``nuclear bomb" is expected to disrupt the entire star while
synthetizing about 0.6 $M_\odot$ (Ia) 
of radioactive $^{56}$Ni, which will power the SN optical light
curves. 
SNIa are very luminous objects and form a quite homogeneous class of
SNe, both in their maximum brightness and their time 
evolution. Thus, SNIa constitute ideal ``standard candles" for distance
determinations on cosmological scales (see Sect. 2.4).

Type Ib/c, on the other hand, must be significantly more massive because
they are only found in spiral galaxies, and often associated with their
spiral arms: this suggests progenitor masses in excess of 5$M_\odot$.
Therefore, either they represent the upper end of the SNIa class or they
are a subclass of core collapse supernovae, possibly massive stars that
occur in binary systems and are able to shed most of their outer H-rich
layers before undergoing the explosion. 


\subsection{Radio Properties}

A series of papers published over the past 18 years on radio supernovae
(RSNe) has established the radio detection and/or radio evolution for 
25 objects: 2 Type Ib supernovae, 5 Type Ic supernovae, and 18 Type II
supernovae.  A much larger list of almost 80 more SNe have low radio
upper limits (\eg Weiler \etal~ 1986, 1998). A summary of the radio
information can be found at: {\it
http://rsd-www.nrl.navy.mil/7214/weiler/sne-home.html}.

All known RSNe appear to share common properties of: 1) non-thermal
synchrotron emission with high brightness temperature; 2) a decrease
in absorption with time, resulting in a smooth, rapid turn-on first at
shorter wavelengths and later at longer wavelengths; 3) a power-law
decline of the flux density with time at each wavelength after maximum
flux density (optical depth $\approx 1$) is reached at that
wavelength; and 4) a final, asymptotic approach of spectral index
$\alpha$ to an optically thin, non-thermal, constant negative value. 

The current model for radio supernovae includes acceleration of 
relativistic electrons and compression of the magnetic field,
necessary for synchrotron emission. These processes occur at the SN
shock interface with a relatively high-density circumstellar medium
(CSM) which has been ionized and heated by the initial UV/X-ray flash
Chevalier (1982a,b). This CSM, which is also the source of the initial
absorption, is presumed to have been established by a constant
mass-loss ($\dot M$) rate, constant velocity ($w$) wind (i.e., $\rho
\propto r^{-2}$) from a red supergiant (RSG) progenitor or a binary
companion. 

In our extensive study of the radio emission from SNe, several effects
have been noted: 1) Type Ia are not radio emitters to the detection
limit of the VLA\footnote{ The VLA is operated by the NRAO of the AUI
under a cooperative agreement with the NSF.}; 2) Type Ib/c are radio
luminous with steeper spectral indices and a fast turn-on/turn-off,
usually peaking at 6 cm near or before optical maximum; and 3) Type II
show a range of radio luminosities with flatter spectral indices and a
relatively slow turn-on/turn-off. These results lead to the conclusion
that most SNII progenitors were RSGs, SNIb/c result from the explosion
of more compact stars, members of relatively massive binary systems,
and SNIa progenitors had little or no appreciable mass loss before
exploding, excluding scenarios that involve binary systems with red
giant companions. 
In some individual cases, it has also been possible to detect thermal
hydrogen along the line of sight (Montes, Weiler \& Panagia 1997, Chu
\etal~1999), to demonstrate binary properties of the stellar system,
and to show clumpiness of the circumstellar material (\eg Weiler,
Sramek \& Panagia 1990). More speculatively, it may be possible to
provide distance estimates to radio supernovae (Weiler \etal~1998). 

\begin{figure} \centerline{ 
\psfig{figure=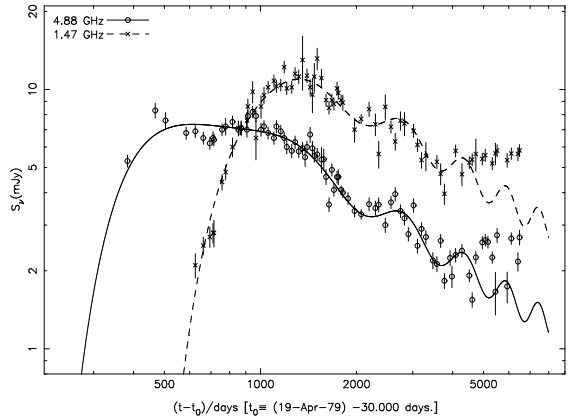,height=6cm}}
\caption{The 1.47 and 4.88 GHz radio emission of SN~1979C as a 
function of time}
\label{fig:rsn79c}
\end{figure}

As an illustration we show that case of SN~1979C that exploded in
April 1979 in the spiral galaxy NGC~4321=M100.  This supernova was
first detected in the radio in early 1980 (Weiler \etal~1981) and is
still bright enough to be accurately measured at different
frequencies, thus offering a unique opportunity to do a very thorough
study of its radio properties, the nature of the radio emission
mechanisms and the late evolution of the SN progenitor. Figure 1
displays the time evolution of SN 1979C radio flux at two frequencies
(1.47 and 4.88 GHz).  One can recognize the ``canonical" properties
(non-thermal spectral index, flux peaking at later times for lower
frequencies, asymptotic power law decline) that allows one to estimate
the circumstellar material distribution, corresponding to a constant
velocity pre-SN wind with a mass loss rate of
$\sim2\times10^{-4}$\msun/year and a probable 20\msun progenitor. In
addition, the almost sinosoidal modulation of the light curves reveals
the presence of a 5\msun binary companion in a slightly elliptical
orbit (Weiler \etal~1992).  And the marked jump up of the flux about
ten years after the explosion (Montes \etal 2000) suggests that the
progenitor had a rather sudden change in its mass loss rate about
10,000 years before exploding, possibly due to pulsational instability
(Bono \& Panagia 1999, in preparation).

\subsection {Supernova Rates} 

Determining the rates of SN explosions in galaxies requires knowing
how many SNe have exploded in a large number of galaxies over the
period of time during which they were monitored.  Although it
sounds easy, this process is rather tricky because data collected
from literature usually do not report the control times over which
the searches were conducted.  On the other hand, more systematic
searches that record all needed information have been started
rather recently and the number of events thus recorded is rather
limited, so that the statistics is still rather uncertain.  In a
recent study, Cappellaro \etal (1999) have thoroughly discussed
this problem and, from the analysis of all combined data set
available, have derived the most reliable SN rates for different
types of galaxies.  We have taken their rates and, for each galaxy
class, we have renormalized them to the appropriate H-band ($\sim
1.65\mu m$) luminosity rather than the B-band ($\sim 0.45\mu m$)
luminosity as done by Cappellaro \etal~(1999). These new rates,
displayed in Table~2, are essentially rates per unit galaxy mass
because the H-band luminosity of a galaxy is roughly proportional
to its mass.  We see that SN rates closely reflect the star
formation activity of the various classes, not only for type II and
Ib/c SNe but also for SNIa.  In particular, the rates for SNII-Ib/c
are 3-4 times higher in late type spirals (Sbc-d) and irregulars
than they are in early type spirals (S0-Sb): this is clear evidence
that star formation is considerably more active in the former than
it is in the latter group. Also, we notice that late type galaxies
(\ie the ones with most active star formation, Sbc through Irr)
have SNIa rates which are 4-10 times higher that the earliest type
galaxies (\ie E-S0). This is a new result (Panagia 1999, in
preparation) and implies that SNIa progenitors are intermediate
mass stars (say, $8>M/M_\odot>3$) and that early type galaxies are
likely to capture and accrete star forming galaxies on a time scale
of one to few billion years to replenish their reservoir of SNIa
progenitors. 

Recent estimates of
the global history of star formation in the Universe were used by
Madau, Della Valle \& Panagia (1998) to compute the theoretical Type
Ia and Type II SN rates as a function of cosmic time from the present
epoch to high redshifts. They show that accurate measurements of the
frequency of SN events already in the range $0<z<1$, and even more so
at higher redshifts, will be valuable probes of the nature of Type Ia
progenitors and the evolution of the stellar birthrate in the
Universe. 

\begin{table}[t]   
\caption{SUPERNOVA RATES {\it (in units of SN per century \& per 
10$^{10}$ L$^H_\odot$)}
\label{tab:rates}}
\vspace{0.2cm}
\begin{center}
\footnotesize
\begin{tabular}{|l|ccc|c|}
\hline
& & & & \\
Galaxy Type 	& SNIa 	   & SNIb/c    & SNII	&  All SNe  \\
& & & & \\
\hline
& & & & \\
E-S0	& 0.05$\pm$0.02	&   $<0.01$     &    $<0.02$	 & 0.05$\pm$0.02  \\
S0a-Sb	& 0.10$\pm$0.04	& 0.06$\pm$0.03 & 0.24$\pm$0.111 & 0.40$\pm$0.12  \\
Sbc-Sd	& 0.21$\pm$0.08	& 0.14$\pm$0.07 & 0.86$\pm$0.35  & 1.21$\pm$0.37  \\
Sm, Irr & 0.59$\pm$0.24 & 0.33$\pm$0.24 & 0.97$\pm$0.60  & 1.87$\pm$0.67  \\
& & & & \\
\hline
\end{tabular}
\end{center}
\end{table}

\subsection{Cosmological Applications} 

As mentioned before, SNIa are virtually ideal standard candles (\eg 
Hamuy \etal~1996) to measure distances of truly distant galaxies,
currently up to redshift around 1 and, considerably more in the
foreseeable future (for a review, see Macchetto and Panagia 1999). 
In particular, Hubble Space Telescope observations of Cepheids in
parent galaxies of SNe Ia (an international project lead by Allan
Sandage) have lead to very accurate determinations of their distances
and the absolute magnitudes of SNIa at maximum light, \ie
$M_B=-19.50\pm0.06$  and $M_V=-19.49\pm0.06$ (\eg Sandage \etal~1996,
Saha \etal~1999). Using these calibrations 
it is possible to determine the distances of much more distant SNe Ia.
A direct comparison with the Hubble diagram (\ie a plot of the
observed magnitudes of SNIa versus their cosmological velocities) of
distant SNe Ia ($30,000~km~s^{-1}>v>3,000~km~s^{-1}$) gives a Hubble
constant (\ie the expansion rate of the local Universe) of
$H_0=60\pm6~km~s^{-1}~Mpc^{-1}$ (Saha \etal~1999). Studying more
distant SNIa (\ie $z>0.1$) it has benn possible to extend our
knowledge to other cosmological parameters. The preliminary results of
two competing teams (Riess \etal~ 1998, Perlmutter \etal~ 1999) agree
in indicating a non-empty inflationary Universe with parameters lying
along the line $0.8\Omega_M-0.6\Omega_\Lambda=-0.2\pm0.1$.
Correspondingly, the age of the Universe can be bracketed within the
interval 12.3--15.3 Gyrs to a 99.7\% confidence level (Perlmutter 
\etal~1999).

\section{Supernova 1987A in the Large Magellanic Cloud}

\subsection{The Early Story}

Supernova 1987A was discovered on February 24, 1987 in the nearby,
irregular galaxy, the Large Magellanic Cloud, which is located in the
southern sky.  SN~1987A is the first supernova to reach naked eye
visibility after the one studied by Kepler in 1604 AD and is
undoubtedly the supernova event best studied ever by the astronomers. 
Actually, despite the fact that SN~1987A  has been more than hundred
times fainter than its illustrious predecessors in the last
millennium, it has been observed in such a detail and with such an
accuracy that we can define this event as a {\it first} under many
aspects (e.g. neutrino flux, identification of its progenitor, gamma
ray flux) and in any case as the {\it best} of all. Reviews of both
early and more recent observations and their implications can be found
in Arnett \etal~ (1989) and Gilmozzi and Panagia (1999), 
respectively.

SN~1987A early evolution has been highly unusual and completely at variance
with the {\it wisest} expectations. It brightened much faster
than any other known supernova: in about one day it jumped from 12th
up to 5th magnitude at optical wavelengths, corresponding to an
increase of about a factor of thousand in luminosity.  However,
equally soon its rise leveled off and took a much slower pace
indicating that this supernova would have never reached those peaks in
luminosity as the astronomers were expecting.  Similarly, in the
ultraviolet, the flux initially was very high, even higher than in the
optical.  But since the very first observation, made with the
International Ultraviolet Explorer (IUE in short) satellite less than
fourteen hours after the discovery, the ultraviolet flux declined very
quickly, by almost a factor of ten per day for several days.  It
looked as if it was going to be a quite disappointing event and, for
sure, quite peculiar, thus not suited to provide any useful
information 	about the other	more common types of supernova
explosions.  But, fortunately, this proved not to be true and soon it
became apparent that SN~1987A  is the most valuable mean to test our
ideas and theories about the explosion of supernovae. 

And even particle emission was directly measured from Earth: on
February 23, around 7:36 Greenwich time, the neutrino telescope
("Kamiokande II", a big cylindrical ``tub" of water, 16~m in
diameter and 17~m in height, containing about 3300 m$^3$ of
water, located in the Kamioka mine in Japan, about 1000~m 
underground) recorded the arrival of 9 neutrinos within an interval of
2 seconds and 3 more 9 to 13 seconds after the first one.
Simultaneously, the same event was revealed by the IMB detector
(located in the Morton-Thiokol salt mine near Faiport, Ohio) and by the 
``Baksan" neutrino telescope (located in the North Caucasus
Mountains, under Mount Andyrchi) which recorded 8 and 5 neutrinos, 
respectively,  within few seconds from each other.  This makes a total of 25
neutrinos from an explosion that allegedly produces 10 billions of
billions of billions of billions of billions of billions of them! But
a little more than two dozens neutrinos was more than enough to verify
and confirm the theoretical predictions made for the core collapse of
a massive star (\eg Arnett \etal~ 1989 and references therein). This
process was believed to be the cause of the explosion of massive stars
at the end of their lives, and SN 1987A provided the experimental
proof that the theoretical model was sound and correct, promoting it
from a nice theory to the description of the truth. 

\subsection {SN 1987A Progenitor Star} 

From both the presence of hydrogen in the ejected matter and the
conspicuous flux of neutrinos, it was clear that the star which had
exploded was quite massive, about twenty times more than our Sun. And
all of the disappointing peculiarities were due to the fact that just
before the explosion the supernova progenitor was a blue supergiant
star instead of being a red supergiant as common wisdom was
predicting. There is no doubt about this explanation because SN~1987A
is exactly at the same position as that of a well known blue
supergiant, Sk~$-69^{\circ}$~202. And the IUE indicated that such a
star was not shining any more after the explosion: the blue supergiant
had gone BANG (Gilmozzi \etal~1987, Kirshner \etal~1987).

On the other hand, common wisdom cannot be wrong and it was {\it not}
quite wrong, after all. At later times, in late May 1987, the IUE
revealed the presence of emission lines of nitrogen, oxygen, carbon
and helium in the ultraviolet spectrum.  They kept increasing in
intensity with time and proved to be quite narrow, indicating that the
emitting matter was moving at much lower speeds (less than a factor of
hundred slower) than the supernova ejecta.  The chemical abundances
and the slow motion were clear sign that that was matter ejected by a
{\it red} supergiant in the form of a gentle wind.  But	there was no
such a star in sight just before the explosion. Therefore, the same
star that exploded, had also been a red supergiant, less than hundred
thousand years before the explosion itself: a short time in the
history of the star but quite enough to make all the difference. 

\subsection{Explosive Nucleosynthesis} 

The optical flux reached a maximum around mid-May, 1987, and declined at a
quick pace until the end of June, 1987, when rather abruptly it slowed down,
setting at a much more gentle decline of about 1\% a day (Pun 
\etal~1995).  Such a decay
has been followed since then quite regularly: a perfect constant decay
with a characteristic time of 114 days, just the same as that of the
radioactive isotope of cobalt, $^{56}$Co, while transforming into iron.
This is the best evidence for the occurrence of nucleosynthesis during
the very explosion: $^{56}$Co is in fact the result of $^{56}$Ni and
this latter can be formed at the high temperatures which occur after the
core collapse of a massive star.  So now, not only are we sure that such
a process is operating in a supernova explosion, just as theorists
predicted, but we can also determine the amount of nickel produced in
the explosion, slightly less than 8/100 of a solar mass or,
approximately, 1\% of the mass of the stellar core before the explosion.  And
the hard X-ray emission detected since July 1987 and the subsequent
detection of gamma-ray emission confirm the reality of this process and
provide more details about its exact occurrence (\eg Arnett \etal~1989
and references therein).

\begin{figure} \centerline{ 
\psfig{figure=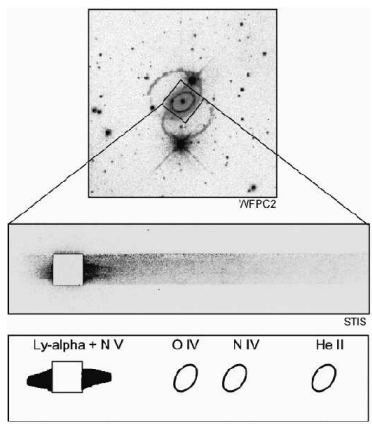,height=7.4cm}}
\caption{The field of view centered on SN~1987A, as viewed by the
HST-WFC2, was observed with HST-STIS  and the resulting ultraviolet 
spectrum is shown in the bottom panels.}
\label{fig:sn87a}
\end{figure}

\subsection{HST Observations} 

The Hubble Space Telescope was not in operation when the supernova exploded,
but it did not miss its opportunity in due time and its first images,
taken with the ESA-FOC in August 23 and 24, 1990, revealed the inner
circumstellar ring in all its ``glory" and detail ({\it cf.} Jakobsen
\etal~1991), showing that, despite spherical aberration,  HST was not a
complete disaster, after all.

Since those early times, Hubble has kept an attentive eye on SN~1987A,
obtaining both imaging and spectrographic observations (\eg Fig. 2) at
least once a year, accumulating valuable data and revealing quite a
number of interesting results (see Gilmozzi \& Panagia 1999), 
such as: \\
- The sequence of images obtained over more than 8 years has allowed
us to measure the expansion of the supernova material
directly: this the first time it has ever been possible and has
permitted to identify the correct models to understand the explosion
phenomenon (Pun \etal~1999, in preparation). \\
- The origin and the nature of the beautiful circumstellar rings are
still partly a mistery.  They have been measured to expand rather
slowly, about 10-20 \kms, \ie 100-2000 times slower than the SN 
ejecta, and to be highly N rich: both these aspects
indicate that the rings were expelled from the progenitor star when it was
a red supergiant, about 20,000 years before the explosion 
(Panagia \etal~ 1996). However, one would have expected such a star to eject
material in a more regular fashion, just pushing away material gently
in all directions rather than puffing rings like a pipe smoker.
Another puzzle is that the star was observed to be a ``blue" supergiant
in the years before the explosion, and not a red supergiant anymore.
This forces one to admit that the star had a rather fast evolution,
which was not predicted by ``standard" stellar evolution theory, and
still is hard to understand fully. \\
- The highest velocity material expelled in SN~1987A explosion has
been detected for the first time by the Space Telescope Imaging
Spectrograph (STIS) (\eg Sonneborn \etal~ 1998). The spectrograph has
found the first direct evidence for material from SN~1987A colliding
with its inner circumstellar ring. The fastest debris, moving at
15,000 \kms are now colliding with the slower moving gas of the inner 
circumstellar ring (Fig. 2) .

In less than a decade the full force of the supernova fast material
will hit the inner ring, heating and exciting its gas and producing a
new series of cosmic fireworks that will offer a spectacular view for
several years. This is going the ``beginning of the end" because in
about another century most, if not all, the material in the rings will
be swept away and disappear, loosing their identities and merging into
the interstellar medium of the Large Magellanic Cloud. This is not a
complete loss, however, because by studying this destructive process,
we will be able to probe the ring material with a detail
and an accuracy which are not possible with current observations.


\section*{References}
\begin{thebibliography}{99}

\bibitem{arn}W.D. Arnett \etal, \Journal{\em ARA\&A}{27}{629}{1989}.

\bibitem{bu}W. Baade and F. Zwicky, \Journal{\it Proc. Nat. Acad. Sci. 
U.S.}{20}{254}{1934}.


\bibitem{cap}E. Cappellaro, R. Evans and M. Turatto, 
\Journal{\AAP}{351}{459}{1999}.

\bibitem{chev82a}R.A. Chevalier, \Journal{\APJ}{259}{302}{1982a}.

\bibitem{chev82b}R.A. Chevalier, \Journal{\APJL}{259}{L85}{1982b}.

\bibitem{chu}Y.H. Chu \etal, \Journal{\APJL}{512}{L51}{1999}.

\bibitem{gil}R. Gilmozzi \etal, \Journal{\em Nature}{328}{318}{1987}.

\bibitem{gipa}R. Gilmozzi and N. Panagia, 
\Journal{\MSAIT}{70}{583}{1999}.

\bibitem{ham}M. Hamuy \etal, \Journal{\AJ}{112}{2391}{1994}.

\bibitem{kir}R.P. Kirshner \etal, \Journal{\APJ}{320}{602}{1987}.

\bibitem{jak}P. Jakobsen \etal, \Journal{\APJ}{369}{L63}{1991}.

\bibitem{mapa}F.D. Macchetto and N. Panagia, in {\it Post-Hipparcos 
Cosmic Candles}, eds. A. Heck \& F. Caputo, p. 225-245 (Kluwer-Holland, 
1999).

\bibitem{mad}P. Madau, M. Della Valle and N. Panagia, 
\Journal{\MNRAS}{297}{L17}{1998}.

\bibitem{}M.J. Montes, K.W. Weiler, \& N. Panagia, 
\Journal{\APJ}{488}{792}{1997}.

\bibitem{}M.J. Montes \etal, \APJ, in press (2000).

\bibitem{pan85}N. Panagia, 1985, in {\em Supernovae As Distance Indicators}, 
Lect.\ Notes Phys.\ Vol.~224, p.~226-240 (Springer-Verlag-Berlin, 1985).

\bibitem{pan91}N. Panagia \etal, \Journal{\APJ}{380}{L23}{1991}.

\bibitem{pan96}N. Panagia \etal, \Journal{\APJ}{457}{604}{1996}.

\bibitem{perl}S. Perlmutter \etal,  \Journal{\APJ}{517}{565}{1999}.

\bibitem{pun95}C.S.J. Pun \etal, \Journal{\APJS}{99}{223}{1995}.

\bibitem{rie}A.G. Riess \etal, \Journal{\APJ}{504}{935}{1998}.

\bibitem{saha}A. Saha \etal, \Journal{\APJ}{522}{802}{1999}.

\bibitem{sand}A. Sandage \etal, \Journal{\APJ}{460}{L15}{1996}.

\bibitem{son}G. Sonneborn \etal, \Journal{\APJ}{492}{139}{1998}.

\bibitem{}K.W. Weiler \etal, \Journal{\APJ}{243}{L151}{1991}.

\bibitem{weil86}K.W. Weiler \etal, \Journal{\APJ}{301}{790}{1986}.

\bibitem{weil90}K.W. Weiler, N. Panagia, \& R.A. Sramek, 
\Journal{\APJ}{364}{611}{1990}.

\bibitem{weil98}K.W. Weiler \etal, \Journal{\APJ}{500}{51}{1998}.


\end{thebibliography}
\end{document}